# GAME THEORETICAL MODELS OF GEOPOLITICAL PROCESSES. PART I.

O.A. Malafeyev,
Saint-Petersburg State University
o.malafeev@spbu.ru

N.D.Redinskikh
Saint-Petersburg State University
redinskich@yandex.ru

V. F. Bogachev
Institute for Regional Economic Studies
Russian Academy of Sciences
vic-bogachev@mail.ru

**Abstract:** In this paper, the interaction of geopolitical actors in the production and sale of military equipment is studied. In section 2 the production of military equipment is considered as the two person zero-sum game. In such game, the strategies of the players are defined by the information state of the actors. The optimal strategy of geopolitical actors is found. In section 3, the conflict process is considered, the optimal strategy is determined for each geopolitical actor.

**Section 1. Preferences**

Let us consider agents purchasing military techniques aviation equipment of types A1, A2, naval equipment of types M1, M2. Agents purchasing military equipment are in groups establishing the preference order of military equipment purchasing.

Table 1.

| Group number | The preference order in the military equipment purchasing |
|---|---|
| 1-st | A1, A2, M1, M2 |
| 2-d | A2, M1, A1, M2 |
| 3-th | M1, A1, M2, A2 |
| 4-th | A1, M2, A2, M1 |

Let us also assume that groups numbered as 1, 2, 3, 4 are with given preference order 5, 15, 35, 45 respectively of the total number of agents purchasing military equipment. Agents that choose another preference order are not presented in Table 1.

Let us construct logical ordering by preference from the priority presented in Table 1 and select best type of military technique. When voting for the best military technique, according to the rule of relative majority, A1 wins with 45%, and the competitor M1 wins with 35%. When comparing submarine with combat helicopter, the greater number of votes belongs to A2: 65% vs. 35%. Next, the M2 wins 85% against 15% for A2. As a result, by comparing M2 and A1, A1 also wins. All voting information is presented in table 2.

By voting for the best military technique, according to the rule of relative majority, A1 wins with 45%, and the nuclear submarine wins with 35%. The greater number of votes belong to the A2 in the comparison with M1: 65% vs. 35%. Next, the A1 gains 85% against 15% A2. As a result, the A1 wins in the comparison M2. All voting information is presented in table 2.

Table 2.

|  | A1 | M1 | A2 | M2 |
|---|---|---|---|---|
| A1 |  | 45% | 80% | 95% |
| M1 | 55% |  | 35% | 50% |
| A2 | 20% | 65% |  | 15% |
| M2 | 5% | 50% | 85% |  |

The given voting method leads to contradictions and does not give any result.

Let us consider another voting method – the Olympic system. Let us divide military equipment into subgroups with different characteristics.

Making use table 2, let us combine military technique into subgroups that does not have common features. Let us consider the following scheme:

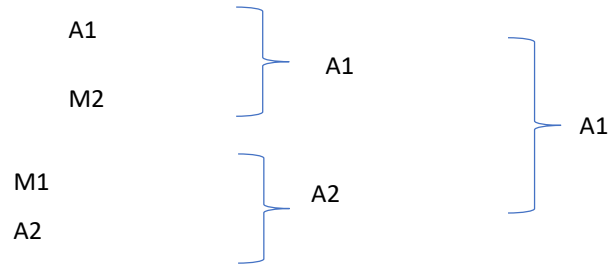

Let us combine military technique into subgroups according to the military power, then the scheme is:

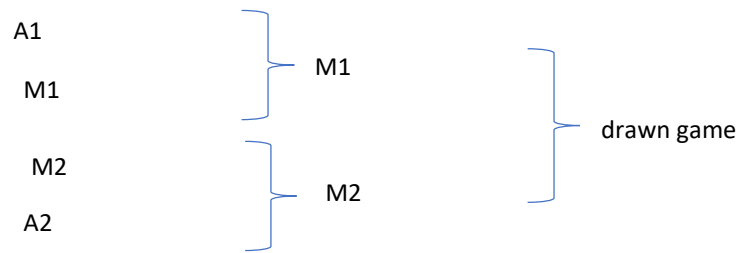

Let us combine military technique by spheres of activity. The A1 wins among the aviation equipment, and there are drawn game among the naval equipment. Then let us compare A1 with M1 and M2 sequentially.

More iteration is for one type technique than for other in such comparison, but this does not affect the result in any way. However, the result depends on the comparison order:

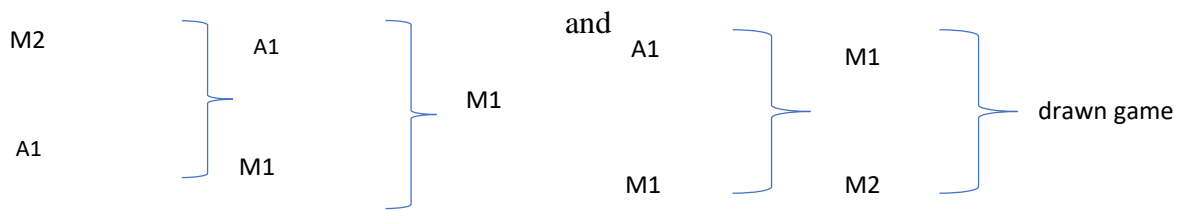

Let us define the best technique making use Table 2. Let us make an analogy with the usual standings of matches between football teams. Here, unlike real matches, the winner objectively chooses based on the points scored. The points scored are in no way related to certain rules of the game. Percentages are the points received during the counter matches. The winner is the one with the highest number of points. Then the points are distributed as follows: 1st place - A1 with 220 points, 2nd – 3rd place – M1 and A1 with 140 points, 4th place - A2 with 100 points. Let us exclude the outsider from our table. Then, with a new calculation, the following result are: A1 with

140 points take the first place, the M1 with 105 points take the 2nd place and M2 with 55 points take the 3rd place. Similarly, let us "cross out" the last. In the "final match", the M1 and A1 meet, where the M1 wins, which contradicts the previous rounds. From the comparisons made by various methods, it is clear that the presented methods may not lead to the final result. Let us consider the concept of "the best", make certain modifications and rules and it is possible to understand which choice of technique is optimal.

Let us introduce the following properties:
1. Internal stability is the property of the prevalence absence of one military equipment over another. That is, after the optimal choice, the opposition of "good to the best" is already excluded.
2. External stability is the property in which for each military equipment not from the optimal, the prevailing technique is found from the optimal.

Such optimal sets with properties of internal and external stability are commonly called Neumann—Morgenstern solutions or N-M solutions. In this case the N-M-solution consists in choosing between naval technique. In this technique internal stability is in the amount of 50% and 50%, and external stability, in the fact that the fighters are opposed by nuclear submarine, and the A2 – M2. There are not other N-M solutions, since there are no pairs with external and internal stability.

### Section 2. Zero-sum game

Let us consider two geopolitical actors plan to produce military equipment. The first geopolitical actor plans to produce aviation equipment, the second geopolitical actor plans to produce only naval equipment. Moreover, each actor can produce only one type of military equipment, which limits the profit of each geopolitical actor. The first geopolitical actor can produce A1 or A2, the second geopolitical actor can produce M1 or M2.

The first geopolitical actor can produce technique of type A1 or A2, the second geopolitical actor can produce technique of type M1 or M2. The cost and market price of all four types of produced military equipment equal. It is assumed, simultaneous production of military technique two types is economically impractical.

It is supposed that produced military equipment is not exported beyond the location of the geopolitical actor. According to experts' forecasts, one thousand units of military technique, of all types can be sold. Two actors distribute the demand for military technique according to the Table 2 in the case of the production.

Let us represent the production of military technique by actors in the form of zero-sum game with two players with constant total payoff. The players in this game are geopolitical actors. The strategies of the players are determined by the information that each actor has. Let us assume that the second geopolitical actor knows what kind of military technique produce the first geopolitical actor. When the first geopolitical actor finds out about the intentions of the second, then it is impossible to change the type of produced military equipment.

Let us describe the strategies in these assumptions. The first geopolitical actor doesn't know about the intentions of the second. It can follow to one of two strategies: produce A1 or A2. In this case, the second geopolitical actor is in one of two information states:
 1) know, that the first geopolitical actor produces A1
 2) know, that the first geopolitical actor produces A2

Accordingly, there are two strategies for each of these states for the second actor:
1) in the response to the production of A1 or A2 to produce M1;
2) in the response to the production of A1 to produce M1, and in the response to the production A2 to produce M2;
3) in the response to the production of A1 or A2 to produce M1
4) in the response to the production of A1 to produce M1, and in the response to the production of A2 to produce M2

5) in the response to the production of A1 to produce M2, and in the response to the production of A2 to produce M1;
6) in the response to the production of A1 to produce M2, in the response to the production of A2 - M2

Let us present the 8 different situations in the table.

Table 3

| Strategy of the second / Strategy of the first | A1 – M1<br>A2 – M1 | A1 – M1<br>A2 – M2 | A1 – M2<br>A2 – M1 | A1 – M2<br>A2 – M2 |
|---|---|---|---|---|
| A1 | A1 – M1<br>45:55 | A1 – M1<br>45:55 | A1 – M2<br>95:5 | A1—M2<br>95:5 |
| A2 | A2 – M1<br>65:35 | A2—M2<br>15:85 | A2—M1<br>65:35 | A2 –M2<br>15:85 |

Let us rewrite Table 3 in the form of matrix, specifying only the first values (since the percentage of equipment sold by the first geopolitical actor determines the percentage sold by the second and gives 100%)

$$\begin{pmatrix} 45 & 45 & 95 & 95 \\ 65 & 15 & 65 & 15 \end{pmatrix} \quad (1)$$

The strategies of the first geopolitical actor are the matrix rows, the strategies of the second geopolitical actor are the columns.

Let us found the optimal strategies of the players in a matrix game with payoff matrix (1). If the first geopolitical factor is interested in its guaranteed payoff when choosing row of the matrix, then it assumes that the second geopolitical actor chooses the column where this row has the minimum element.

Let us introduce two definitions:

1. The element in the string that is the largest among the smallest let us call the maximin. The strategy of the player let us call, respectively, maximin.

2. The largest element in the column that is the minimum among the largest let us call minimax, and the strategy, respectively, minimax.

It is easy to explain why these definitions are introduced. It is advantageous for the first geopolitical actor to choose the maximum payoff that it can receive. However, the first geopolitical actor should understand that its payoff is limited, therefore, the first geopolitical actor should choose strategy where its losses from limited by the second are minimal. That is, choose the maximum strategy among the smallest. Let us do the similarly for the second geopolitical actor.

In this case, the first geopolitical actor selects the first row with element 45, and the geopolitical actor 2 selects the second column with element 45.

The maximin strategy of the first geopolitical actor has the following stability property: deviation of the actor from the maximin strategy cannot lead to an increase in its guaranteed payoff. Therefore, the first geopolitical actor has no reason to deviate from the maximin strategy. Any deviation of the second geopolitical actor from the minimax strategy cannot reduce its possible losses.

In the considering game the situation formed by the choice of the first row and the second column of the matrix is such that it is unprofitable for any of the actors to deviate from it. The resulting situation is an equilibrium. Thus, in order to have an equilibrium in zero-sum game, the equality of maximin and minimax in the payoff matrix is necessary and sufficient.

Thus the optimal strategy of the first geopolitical actor is the production military technique of type A1, and the optimal strategy of the second geopolitical actor is the production military technique of type M1.

**Section 3. Conflict process without information. Mixed strategies.**

Let us consider the problem from section 2. For problem condition from the section 2 let us introduce the next assumptions: both geopolitical actors do not know about each other's intentions to produce military technique and have two alternatives: first geopolitical actor can produce aircraft equipment of the type A1 or A2. Second geopolitical actor can produce naval equipment of the type M1 or M2. These alternatives are the strategies of geopolitical actors.

The resulting game is matrix game. This game has 2x2=4 situations, and the payoff matrix has a form:

$$\begin{array}{c} \quad\quad M1 \quad M2 \\ \begin{array}{c} A1 \\ A2 \end{array} \left[ \begin{array}{cc} 45 & 95 \\ 65 & 15 \end{array} \right] \end{array} \quad\quad (2)$$

First geopolitical actor chooses the matrix rows. Second geopolitical actor chooses the matrix columns. The matrix (2) is part of the matrix (1). The equilibrium situation of the previous game does not exist in the matrix (2), since the entire second column of the matrix (1) is missed in the matrix (2). Another equilibrium situation in the matrix (2) does not occur, and the maximin is different from the minimax. The maximin optimality principle is unrealizable in such form as it implements in the game in section 2.

In such game, the first geopolitical actor can get more than 45. In this case, the second geopolitical actor is interested in reducing the capabilities of the first geopolitical actor. In the considering situation, it is not impossible to implement.

Let us extend the strategy areas of the players (geopolitical actors) by adding randomness to the choice of strategies. By randomness, let us understand the ability of players to calculate certain probabilities in advance and make a decision.

To simplify, let us assume that the probabilistic event occurs once.

For example, let us denote by random variable the payoff of the first geopolitical actor. If it chooses the first row with some probability p, then the second one – with probability 1-p. The second geopolitical actor chooses the first column. It follows from matrix (2) that the payoff takes the value 45 with probability p, while the payoff with value 65 is taken with probability 1-p.

For simplicity, let us assume that the dependence is linear and the expected payoff of the first geopolitical actor is:
$$45p+65(1-p). \quad\quad (3)$$

Similarly, if the first geopolitical actor proceeds as above and the second geopolitical actor selects the second column of the matrix (2), then the expected payoff of the first geopolitical actor equals:
$$95p+15(1-p). \quad\quad (4)$$

Let the first geopolitical actor behaves absolutely the same as before, and the second actor decides to introduce randomness into its choice. It chooses the first column with probability q, and the second column with 1-q. It is assumed that the probability q does not change whether the first geopolitical actor chooses its first strategy or the second. The choice of each geopolitical actor is independent. It follows from this that the expected payoff of first actor is:
$$[45p+65(1-p)] q + [95p+15(1-p)] (1-q),$$
or
$$15+80p+50q-100pq \quad\quad (5)$$

Now let us consider new game with first and second geopolitical actors. The strategy of the first actor is in choosing the first row with probability p, $0 \leq p \leq 1$. The strategies of first actor is in choosing non-negative number p, which is not greater than one, while being the probability of choosing the first row of the matrix. The strategies of the second geopolitical actor are in choosing the first column with the probability q. The expression (5) describes the value of payoff function in the situation resulting from the choice of probabilities p and q.

Such random, probabilistic actor strategies are mix strategies. Note that for p=1 or p=0, the strategy of the first geopolitical actor is in reliably choosing the first or, respectively, the second row of the matrix. Thus, all the old strategies are among the new, mixed ones. Such strategies are pure strategies.

Similarly, mixed strategies of the second geopolitical actor are introduced and its pure strategies stand out among them. The mixed strategy of geopolitical actor, consisting in choosing the first pure strategy with probability t, and the second pure strategy with probability 1—t, is denoted as pair of numbers (t, 1-t).

Let us consider the minimaxes of expression (5). Let us found the maximum value of the minimum payoff of the first geopolitical actor. Let us imagine (5) in the form:
$$15+80p+(50—100p)q. \tag{6}$$

For $p \leq \frac{1}{2}$, the coefficient in (6) for q is non-negative and, therefore, the second geopolitical actor chooses the smallest value of q, i.e. 0, to reduce the payoff of the first geopolitical actor. Then the payoff of the first geopolitical actor equals to 15+80p. To maximize this expression, it is advantageous for the first geopolitical actor to take the largest p allowed in this area, i.e. $\frac{1}{2}$. His winnings in this case will be equal $15 + 80 * \frac{1}{2} = 55$.

For $p \geq \frac{1}{2}$, the coefficient in (6) for q is not positive and it is advantageous for the second geopolitical actor to choose q=1. The payoff of the first geopolitical actor equals to 65—20p, and the first geopolitical actor chooses the smallest of the p, where $p \geq \frac{1}{2}$, i.e. the number $\frac{1}{2}$ itself, while receiving an expected payoff of 55. In both cases, the maximum payoff of the actors equals, since the largest values are achieved here at the same value p=$\frac{1}{2}$.

Let us present the above reasoning on the graph.

The value of maximin is 55 with mixed strategy $(\frac{1}{2}, \frac{1}{2})$ of the first geopolitical actor

Now let us found minimax. Transform (5) as follows:
$$15+50q+(80—100q)p,$$
in the case of $q \leq 0.8$, the maximum is reached at p=1 and is equal to:
$$15+50q+80—100q=95—50q,$$
and if $q \geq 0.8$, then the maximum is reached at p=0 and is equal to 15+50q.

Therefore, for the second geopolitical actor, it is advisable to take q = 0.8 (i.e. a mixed strategy $(\frac{8}{10}, \frac{2}{10})$), and the minimax equals 18+50*0,8 =55.

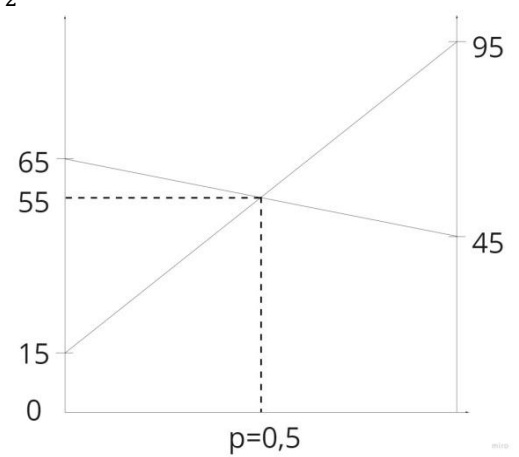

Consequently, the mixed strategies $(\frac{1}{2}, \frac{1}{2})$ of the first geopolitical actor and $(\frac{8}{10}, \frac{2}{10})$ of the second geopolitical actor are their optimal strategies. The number 55 is the value of the game.

Thus, the optimal strategy of the first geopolitical actor is the production of aviation equipment with probabilities of 0.5. The optimal strategy of the second geopolitical actor is the production technique of type M1 with probability of 0.8 and the production of M2 with a probability of 0.2. The expected success of the first is 55%, and the success of the second is 45%.

It should be noted, that if there is information about the plans of the first geopolitical actor, the second geopolitical actor can produce 60% of the equipment of choosing type instead of 45%.